\documentclass{article}
\usepackage{graphicx} 
\usepackage{amsmath}
\usepackage{amssymb}
\usepackage{hyperref, natbib}
\usepackage[textsize=tiny]{todonotes}

\title{On functional freedom and Penrose's critiques of string theory\footnote{Published in \emph{Philosophy of Physics} 2(1), 2024.}}
\author{Matěj Krátký\footnote{University of Geneva, CH. matej.kratky@etu.unige.ch}~ \& James Read\footnote{University of Oxford, UK. james.read@philosophy.ox.ac.uk}}
\date{}

\begin{document}

\maketitle

\begin{abstract}
In his \textit{The Road to Reality} as well as in his \textit{Fashion, Faith and Fantasy}, Roger Penrose criticises string theory and its practitioners from a variety of angles ranging from conceptual, technical, and methodological objections to  sociological observations about the string theoretic scientific community. In this article, we assess Penrose's conceptual/technical objections to string theory, focussing in particular upon those which invoke the notion of `functional freedom'. In general, we do not find these arguments to be successful.
\end{abstract}

\tableofcontents

\section{Introduction} \label{introduction}

In his \textit{The Road to Reality} \cite{RR} as well as in his \textit{Fashion, Faith and Fantasy} \cite{FFF}, Roger Penrose criticises string theory and its practitioners from a variety of angles ranging from conceptual, technical, and methodological objections to basically sociological observations about the string theoretic scientific community. In this article, we assess Penrose's conceptual/technical objections to string theory, focussing in particular upon those which invoke the notion of `functional freedom'.

Roughly speaking, for a fairly arbitrary field theory formulated on some smooth manifold, the notion of functional freedom seeks to quantify the degrees of freedom associated with the given field. Intuitively, one might be tempted to say that there is in some yet-to-be-specified sense \textit{more} freedom associated with a vector field in Minkowski spacetime than with a mere scalar field because a vector is a multi-component object: in order to specify fully a vector in tangent space one in general needs to specify more than just one real number. A scalar field, on the other hand, is specified fully by giving only a single number at every point of the manifold. Functional freedom attempts to cast this rough idea into more perspicuous and mathematically rigorous terms.

In \S\ref{FF}, we attempt to give the most charitable reconstruction of the notion of functional freedom. We first review how Penrose  introduces the concept heuristically in his books, before proposing a formalization of functional freedom in terms of constrained Hamiltonian systems. While this provides a rigorous footing for the heuristics, we ultimately conclude that the scope of applicability of this formalization is limited only to a narrow class of theories. We take this to be problematic because we read Penrose as proposing functional freedom as an equivalence criterion for physical theories such as string theory which \textit{do not} fall into this category. Separately from this, we also provide an independent evaluation of Penrose's proposal of using functional freedom as an criterion of theoretical equivalence.
After a brief introduction to perturbative bosonic string theory in \S\ref{string theory}, in \S\ref{arguments} we then reconstruct, summarize, and evaluate critically Penrose's specific arguments which invoke the notion of functional freedom.
In \S\ref{conclusion} we conclude.

\section{Functional Freedom} \label{FF}
Physics is replete with field theories in various guises. Some paradigmatic examples include classical electromagnetism, fluid mechanics, Newton-Poisson gravity, the general theory of relativity, analytical mechanics, and quantum field theory.

Let's attempt to be more precise about what's meant by a `field theory'. In each of the above cases, one starts by considering a smooth manifold $M$, which can be thought of as a kind of background on which a physical story unfolds. The physical story concerns some additional degrees of freedom called fields, which assign an attribute or some value to each point in $M$. If one labels the fields by $\phi_i$, one can think of them as functions $\phi_i: M \rightarrow F_i$, where $F_i$ is a differentiable manifold of field values. On the semantic view of scientific theories, the kinematically possible models (KPMs) of a generic field theory are thus specified by ordered tuples $\langle M, \phi_1, ..., \phi_n \rangle$. The dynamically possible models (DPMs) are then a subset of KPMs picked out by appropriate dynamics, e.g.~the Einstein field equations or Maxwell's equations, or perhaps path integrals in the case of quantum field theories. A field theory thus consists of a specification of the KPMs together with some dynamics which allow one to pick out the DPMs from amongst the KPMs.

We have alluded above to the fact that functional freedom (henceforth FF)---discussed by Penrose at \cite[pp.~378--80]{RR} and \cite[pp.~401--7]{FFF}---can be understood intuitively as a means by which to quantify and compare the number of degrees of freedom present in field theories. In \S\ref{Penrose on FF}, we introduce the concept as conceived by Penrose, quoting the relevant passages from the above texts; in \S\ref{conceptual eng}, we assess Penrose's efforts through the lens of conceptual engineering. We think that there are good reasons to believe that FF may in fact be understood as an attempt at revision of the mathematical concept of quantity. Then, in \S\ref{Sec Constraints} we propose a formalization of FF using constrained Hamiltonian systems, and in \S\ref{equivalence} we consider the relations between functional freedom and notions of theoretical equivalence. 

\subsection{Penrose on Functional Freedom}\label{Penrose on FF}
In a generic field theory such as classical electromagnetism, the number of possible field configurations over a background manifold is  infinite. More precisely, the cardinality of the set of KPMs will be $2^{\aleph_0}$ regardless of the precise dimension of $M$ or of the $F_i$.
We don't provide a rigorous proof of this statement but a reflection on the various examples mentioned above hopefully lends it at least some plausibility. Furthermore, \cite[pp.~378--80]{RR} contains some discussion by Penrose of the various sizes of infinity encountered in physics with the conclusion that most such physical sets (manifolds, sets of continuous functions on manifolds, Hilbert spaces, possible particle trajectories) are of cardinality  $2^{\aleph_0}$. But given that this cardinality is the same for all such cases, if one is to quantify and compare the degrees of freedom between field theories, how is one to proceed?

Penrose's texts offer some intuitions. Let's start by approximating $M$ and $F_i$ by discrete, compact manifolds consisting of only a finite number of points. Note that one might try to recover the original manifolds from this approximate picture by taking the limit as the number and density of points grows to infinity. Subsequently, one might proceed by counting combinatorially the number of possible field configurations since discretization reduced the infinity to a finite number. We illustrate this general procedure with a specific example from \textit{Fashion, Faith and Fantasy} \cite{FFF}.\footnote{Note a related attempt to cash out functional freedom in terms of fibre bundles at \cite[pp.~439--45]{RR}.}

Consider the following simple field theory $\mathcal{T}$ whose KPMs are of the form $\langle \mathbb{R}^3, v \rangle$, where $v$ is a three-dimensional vector field on $\mathbb{R}^3$. For simplicity, we don't assume any dynamics for $\mathcal{T}$ so (in some sense) all the KPMs are also the DPMs. One may try to approximate $\mathcal{T}$ by the following a discretized model following \cite[pp.~401--4]{FFF}. Consider a three-dimensional cubic lattice of $N^3$ points (formerly $\mathbb{R}^3$) on which there lives a discretized three-vector field (formerly $v$). In other words, suppose that each of the $N^3$ points is assigned a vector whose components may take one of $K$ values. Note that $N, K \in \mathbb{N}$. A simple combinatorial argument then yields $(K^3)^{3N} = K^{3N^3}$ as the total number of possible field configurations on the lattice since each of the $N^3$ points can be assigned one of $K^3$ possible vectors. Extending this construction to a more general case, one finds that for a field with $c$ independent components over a $d$ dimensional lattice, the same argument would give the functional freedom of $K^{cN^d}$. In the continuum limit, $K^{cN^d}$ becomes infinite and the above analysis seems to break down. However, in trying to retain the sense of magnitude captured elusively by the above analysis, Penrose opts for a novel notation. He introduces the formal expression $\infty^{c\infty^d}$ to denote the \textit{functional freedom} of $\mathcal{T}$.\footnote{Penrose is explicit that he owes this notation to Wheeler---see \cite{Wheeler}.}

Suppose now that $\mathcal{T}$ does have non-trivial dynamics. Intuitively, dynamical equations should decrease the functional freedom since the fields are now constrained by the dynamical equations. To capture this fact, we introduce the following distinction. We call \textit{kinematical functional freedom} (KFF) the functional freedom of a theory obtained by ignoring its dynamics. We call \textit{dynamical functional freedom} (DFF) the functional freedom which takes the full dynamics into account. Ultimately, DFF rather than KFF will be at the centre of our attention in the subsequent discussion, but (alas) determining DFF of theories with non-trivial dynamics is a delicate issue.\footnote{This is related to the fact that, for example, to identify the full set of symmetries of some dynamics is a highly non-trivial task.}

Let us demonstrate the procedure with the familiar example of vacuum Maxwell theory in four dimensions whose KPMs are the triples $\langle M, \eta, F \rangle$, with $M$ a smooth manifold, $\eta$ a Minkowski metric on $M$, and $F$ a two-form. It turns out that for the purposes of obtaining DFF, it is actually convenient to represent this theory non-covariantly in terms of three-component electric and magnetic field vectors $\Vec{E}$ and $\Vec{B}$. The
KFF of the theory is $\infty^{6\infty^4}$ since one must populate four-dimensional spacetime with two three-component vector fields. Dynamics is implemented via the Maxwell equations
\begin{align}
    \nabla \cdot \Vec{E} &= 0, \label{max1} \\
    \nabla \cdot \Vec{B} &= 0, \label{max2} \\
    \nabla \times \Vec{E} &= -\frac{\partial\Vec{B}} {\partial t}, \label{max3} \\
    \nabla \times \Vec{B} &= \frac{\partial \Vec{E}}{\partial t}. \label{max4}
\end{align}
Importantly, the above equations admit a convenient initial value formulation. Consider a three-dimensional spacelike hypersurface $\Sigma$ in $M$ which, for simplicity, may be taken to be the flat Minkowski spacetime isomorphic to $\mathbb{R}^4$. Once the values of $\Vec{E}$ and $\Vec{B}$ are specified on $\Sigma$, one can use (\ref{max3}) and (\ref{max4}) to evolve the fields to the future of $\Sigma$. This would seem to decrease the FF from the original KFF of $\infty^{6\infty^4}$ to $\infty^{6\infty^3}$ since the equations ensure that to completely fix the DPMs one need only populate the three-dimensional $\Sigma$ with field values. However, one must exercise further care when specifying $\Vec{E}$ and $\Vec{B}$ on $\Sigma$. In particular, the two constraints (\ref{max1}) and (\ref{max2}) further decrease the freedom to specify the fields on $\Sigma$. As a result, the FF further decreases to $\infty^{4\infty^3}$ where one degree of freedom is subtracted for each constraint equation. This is the DFF of vacuum Maxwell theory in four-dimensions, in agreement with \cite[p. 186]{Penrose_2002}. Heuristically speaking, one therefore needs to subtract one from the lower exponent for each constraint equation and one from the upper exponent thanks to the existence of the initial value formulation. We recognize that this example can serve only as a mere heuristic and in \S\ref{Sec Constraints} we discuss how the counting procedure can be made more rigorous.


Unfortunately, the above examples and other similar such examples in \cite{RR, FFF} seem to exhaust the discussion of the concept of FF present in Penrose's texts. Most regrettably, very little or none of the attention is devoted to the apparent clash in linguistic practice which occurs when one compares FFs to each other and refers to them as if they were regular magnitudes. To this end, Penrose insists that FF does capture a sense of magnitude of the set of DPMs or of its respective field theory (note: Penrose does not use the terminology of `DPMs'), and yet resists stringently its identification with Cantor's theory of infinite cardinals. We quote at length:
\begin{quote}
    The first point I should make is that these infinite numbers do not refer to the ordinary (Cantor) sense of cardinality that describes the sizes of general infinite sets. [...] Cantor’s theory of (cardinal) infinities is really concerned just with sets, which are not thought of as being structured as some kind of continuous space. For our purposes here, we do need to take into account continuity (or smoothness) aspects of the spaces that we are concerned with. For example, points of the 1-dimensional line $\mathbb{R}$ are just as numerous, in Cantor’s sense, as the points of the 2-dimensional plane $\mathbb{R}^2$ (coordinatized by the pairs $x$, $y$ of real numbers) [...]. However, when we think of the points of the real line $\mathbb{R}$ or the real plane $\mathbb{R}^2$, respectively, as organized into a continuous line or a continuous plane, the latter must indeed be thought of as much “larger” entity, in the limit when the finite $N$-element set $\textbf{R}$ becomes continuous $\mathbb{R}$. \cite[pp.~405--7]{FFF}
\end{quote}
We take it to be manifest in the above that Penrose indeed takes expressions like $\infty^{c\infty^d}$ to capture a meaningful sense of magnitude. This is made evident further by his very explicit comparisons of such quantities. For example, in the texts one finds arguments to the effect that when one compares the functional freedom of two theories, it is the dimension of the background manifold which effectively determines which functional freedom is larger.
This point is again justified by an analogy with the finite cases where the inequality $K^{C N^{D}} \gg K^{c N^{d}}$ holds if $D>d$, the values of $C$ and $c$ being effectively irrelevant. In the continuum limit, one then obtains the formal inequality $\infty^{C\infty^{D}} \gg \infty^{c\infty^{d}}$ capturing the relative magnitudes of the two functional freedoms. One is told that:
\begin{quote}
The double inequality sign “$\gg$” is used in order to convey the utter unassailable hugeness whereby the functional freedom described by the left-hand side exceeds that described by the right-hand side, when the spatial dimensionality is greater, no matter what the component numbers $C$ and $c$ are  [...] Such a theory cannot be \textit{equivalent} to another such theory on which the initial space has a different number $D$ of dimensions. If $D$ is greater than $d$, then the freedom in the $D$-space theory always vastly exceeds that in the $d$-space theory! \cite[p.~41, emphasis added]{FFF}
\end{quote}
Importantly, note that in the above, Penrose sees FF as a criterion on equivalence of theories. We will pick up this theme in \S\ref{equivalence}, but before doing so, here is yet another example which can be used to fire up intuitions about FF. Consider elements of $\mathbb{N}^2$ arranged in a two-dimensional lattice. One can then follow Cantor's diagonal argument and trace out the entire lattice with a `squiggly' line to prove graphically that $\mathbb{N}^2$ is denumerable and has the same cardinality as $\mathbb{N}$. However, as per the above quote, Penrose would like to claim in some sense $\mathbb{N}^2$ is \textit{larger} than $\mathbb{N}$. He makes the following observation:
\begin{quote}
    [That $\mathbb{N}^2$ is larger than $\mathbb{N}$] is illustrated by the fact that the counting procedure for pairs [i.e.,~Cantor's argument] cannot be made to be continuous. (Though ``continuous'' in the limited sense that ``close'' elements of our counting sequence indeed always give us ``close'' pairs $(r, s)$, it is not true in the necessary technical reverse sense that close pairs always give close members of the counting sequence.) \cite[pp.~405--7]{FFF}
\end{quote}
To reiterate, the suggestion seems to be that the existence or absence of continuous maps between the two sets should capture the sense of magnitude and subsequent comparisons. Whether existence of continuous maps could provide a good criterion to capture FF (and perhaps a definition) is a mathematical question which for now we will not attempt to answer; however, the example serves as a further datum for Penrose's intuitions regarding FF.

On more general grounds, it seems to us that one should be worried about the soundness of comparisons of FF and about the apparent clash in linguistic terminology between FFs and Cantor's cardinals when both, seemingly, are used to perform comparisons of magnitude. If Penrose maintains that functional freedom is \textit{not} a cardinal number, and yet believes that two functional freedoms may meaningfully be compared in size, then it is tempting to simply accuse him of inconsistency. After all, official linguistic practice in mathematics understands terms `larger', `smaller', `more', and `less' in terms of Cantor's cardinals or (alternatively) in terms of measure theory. Engaging with this practice by making comparisons of functional freedom would therefore seem to commit Penrose to FF just being Cantor's cardinals, which is exactly what he denies.\footnote{Note that there are indeed good reasons for Penrose to deny this since if FFs \textit{are} understood as Cantor's cardinals, then all their comparisons are rendered trivial as per our initial discussion of infinities encountered in physics. Briefly, for any set $S$, such that $\lvert S \rvert \geq \aleph_0$, one has $\vert S \rvert ^n = \lvert S \rvert$ for any $n \in \mathbb{N}$. In other words, exponentiation of infinite cardinals by a natural number doesn't change their cardinality. This is particularly relevant to finite dimensional differential manifolds in which we take interest in physics. The cardinality of such manifolds is  $2^{\aleph_0}$ regardless of their dimension which means that all functional freedoms, if understood as cardinal numbers, are the very same cardinal number, namely ${\left(2^{\aleph_0} \right)}^{2^{\aleph_0}}$, or ${\left(2^{\aleph_0} \right)}^{\aleph_0}$ for continuous functions which are defined by their values on rational points.}

So it seems that Penrose bears some further burden of explication. If FFs are not cardinals and yet can meaningfully be compared, then how exactly is this to be cashed out?\footnote{Brian Pitts has suggested to us that some recent philosophical work on `how to measure the infinite' might be put to work in order to help Penrose out on these issues---see in particular e.g.\ \cite{BenciDiNassoInfiniteBook}. However, since that work applies only to countable infinities, it is not obviously applicable to the field theory contexts which are relevant here.} What justifies drifting away from the official linguistic practice of mathematics? We suggest that Penrose might plausibly be understood as attempting to \textit{revise} the mathematical concept of magnitude. We devote the following subsection to this line of thinking, invoking some recent work on the topic of conceptual engineering.

\subsection{Conceptual Engineering} \label{conceptual eng}

At this point, we would like to make a connection with recent philosophical literature on conceptual analysis and concept engineering \cite{Cappelen}. We propose that what Penrose seems to be doing in the above passages is instigating an attempt to \textit{revise} the accepted mathematical conceptual vocabulary. As Cappelen reminds us, revision of concepts is an ubiquitous practice in philosophy and other areas of human intellectual endeavour, so this observation perhaps doesn't come as a great surprise. Nevertheless, we take it to be a sign of good philosophical practice to be explicit about such moves in order for an evaluation of the proposed conceptual revision to be available.

A brief survey of examples of conceptual engineering in philosophy and beyond can be found in \cite{Cappelen}. To illustrate the point, we mention here just one example of what we take to be a successful instance of conceptual engineering stemming from the metaphysics of race. There are at least three metaphysical positions about race, reviewed nicely in \cite[ch.~10]{Ney}. These positions regard race as either (i) a social kind, (ii) a biological, natural kind, or (iii) an illusory non-existent category. Following Ney, we refer to these views as \textit{social constructivism}, \textit{biological realism}, and \textit{eliminativism}, respectively. While debates are still ongoing, we believe that sufficient philosophical, scientific, and political consensus has been reached on this matter and races are now understood either in spirit of social constructivism or eliminativism rather than biological realism. One can rightly regard this as an instance of conceptual engineering since historically race was often regarded as a biological category.


We confessed already in \S\ref{Penrose on FF} that Penrose's heuristic examples don't suffice to make DFF fully rigorous; however, we believe that DFF can be made such using the machinery of Hamiltonian dynamics, to be discussed in \S\ref{Sec Constraints}.
That said, even if this approach to making DFF rigorous is indeed to be undertaken, it doesn't remove the burden explicating and arguing for the relevance and general usefulness of such a notion. For it is one thing to (i) meaningfully define a concept, another thing to (ii) demonstrate its practical usefulness in applications, and yet another thing to (iii) demonstrate its superiority over other linguistically entrenched concepts and propose their revision in light of such findings. In the above passages, we read Penrose as engaging most actively with (i) and (iii). Engagement with (i) in Penrose's texts seems indubitable due to his multiple examples which might legitimately be regarded as definition of a concept via ostension. 
Furthermore, frequent reference to comparisons of \textit{size} of FFs suggests strongly Penrose's engagement with (iii) since the linguistic territory previously reserved for the established mathematics of Cantor's cardinals might rightly be regarded as invaded by the new concept of FF. Regardless of whether we read Penrose as merely making a call for \textit{disambiguation} or alternatively for a thorough \textit{revision} of our concepts, it is precisely this fact which makes FF interesting philosophically. With regards to (ii), we prefer to read Penrose as finding use for the concept of DFF in a kind of equivalence criterion for theories. We formulate and evaluate this criterion in \S\ref{equivalence}.

Speaking much more broadly, one can ask the following questions about conceptual engineering and revision: what makes a revision of a concept desirable? What makes it successful? When is it appropriate to define a new concept? Below, we propose a defeasible list of ways in which new concepts can be evaluated; however, we note that a proper engagement with the above questions and relatedly with the criteria for successful conceptual engineering and concept revision are well beyond the scope of our current project:
\begin{itemize}
    \item \textit{Coherence}. A concept fails to be coherent if its application leads in one way or another to a contradiction. For example, the composite concept of being short and tall at the same time is incoherent because the properties of being short and being tall arguably cannot be true of a single individual at the same time. Perhaps more controversially,  van Inwagen famously proclaimed the concept of free will to be incoherent: see \cite[p.~17]{Cappelen}.
    \item \textit{Metaphysical credibility}. A concept is metaphysically credible just in case it latches well onto salient features of reality.\footnote{Compare the distinction between sparse and abundant properties due to  Lewis \cite{Lewis}.} 
    \item \textit{Usefulness}. Concepts may be useful in many ways. They may allow us to talk about certain tasks meaningfully, via their connectedness to other concepts, via their explanatory power, etc.
    \item \textit{Connectedness} of a concept captures how well the new concept fits into the broader system of concepts which one has already adapted.
\end{itemize}

We'll leave things here. We now introduce what we take to be a rigorous way to count FF and then turn to Penrose's suggestion that FF can be useful in adjudications on theoretical equivalence, which is relevant to (ii) above.


\subsection{Functional Freedom from Constraint Analysis}\label{Sec Constraints}
While the discussion in \S\ref{Penrose on FF} reviews Penrose's writing on the topic of counting DFF to our best knowledge exhaustively, we believe that counting DFF can be rendered mathematically rigorous by using the formalism of constrained Hamiltonian systems (see e.g.\ \cite{Henneaux:1992ig}). It should be noted that while Penrose's (implicit) commitment to this counting method is quite probable given the results presented in his texts, we are not aware of Penrose's explicit commitment to it. Nevertheless, we take it to be the most charitable reconstruction of Penrose's heuristic DFF counts and we believe we do no harm to Penrose in reading his texts this way.
We spend the rest of this subsection explaining how one can  define rigorously the DFF of a field theory which admits a Hamiltonian formulation, recovering the $\infty^{4\infty^3}$ result for classical electromagnetism already discussed above. By restricting attention to field theories which admit of a Hamiltonian description, we of course limit significantly the scope of the inquiry---we'll return to this below. In order for a field theory to admit Hamiltonian formulation, one needs to privilege a `time' parameter, foliate $M$ in to constant time slices, and be able to view the dynamics of the theory as describing a time-evolution of fixed time configurations. This will generally not be the case: GR, for instance, admits such a $(3+1)$-formulation only in certain specific cases (see e.g.\ \cite[ch.~10]{Wald1984} for more detail).

Suppose now that that our field theory admits a Hamiltonian formulation and that it is described by the Lagrangian function $\mathcal{L}$ on some $2n$-dimensional tangent bundle $T\mathbb{Q}$ with local coordinates $(q^1,\dots,q^n,\dot{q}^1,\dots,\dot{q}^n)$. To obtain the Hamiltonian, one must perform a Legendre transform of the Lagrangian which amounts to taking $\mathcal{H}=p_i\dot{q}^i -\mathcal{L}$, and expressing $\mathcal{H}$ entirely in terms of $q^i$ and $p_i$, where $p_i=\frac{\partial\mathcal{L}}{\partial\dot{q}^i}$ are the generalized momenta. As it turns out, in certain cases this is not possible because the relations involving generalized momenta fail to be invertible and instead define a \emph{constraint}.

It is a well-known fact that the constraints admit the following twofold classification. Firstly, we distinguish whether the constraint is enforced as a matter of kinematics or dynamics:
\begin{itemize}
    \item \textit{Primary constraints} are enforced kinematically and stems directly from the non-invertibility of the $p_i=\frac{\partial\mathcal{L}}{\partial\dot{q}^i}$ relations.
    \item \textit{Secondary constraints} are enforced as a matter of dynamics and stem from the requirement that primary constraints be preserved under Hamiltonian evolution.
\end{itemize}
Having collected all primary and secondary constraints in a single batch $\mathcal{C}$, we next distinguish between
\begin{itemize}
    \item \textit{First-class constraints} which, roughly speaking, are the maximal subset $\mathcal{F}\subseteq\mathcal{C}$ such that the Poisson brackets of any two constraints in the subset vanishes on the constraint surface. In other words, the subset forms an algebra under the Poisson bracket.
    \item \textit{Second-class constraints} which occur whenever $\mathcal{C}$ does not close to an algebra. We denote them by $\mathcal{S}\subseteq\mathcal{C}$.
\end{itemize}
First-class constraints are usually in a naïve way thought of as generators of \textit{gauge transformations}.\footnote{Note that this common wisdom has been recently subject to debate in \cite{Pitts2014,pooley2022firstclass}.} According to Henneaux and Teitelboim \cite[p.~29]{Henneaux:1992ig}, one can then define the number of degrees of freedom $c$ of the theory under consideration as follows:
\begin{equation}\label{Eqn DOFcount}
    2c = 2n - |\mathcal{S}| - 2|\mathcal{F}|.
\end{equation}

The central claim of this section is that the number of degrees of freedom (DOFs) $c$ is in fact directly related to the lower exponent in Penrose's FF notation. To see that this agrees with our previous results, consider again the example of electromagnetism in Minkowski spacetime. The configuration space $\mathbb{Q}$ in this case consists of tuples $(\Vec{A},\Vec{E})$, where $\Vec{A}$ is the vector potential and $\Vec{E}$ the electric field. In Penrose's notation $\mathbb{Q}$ is therefore $\infty^{6}$ dimensional. In the Hamiltonian formulation, electromagnetism further has one primary and one secondary constraint, both of which are first class.\footnote{This is the vanishing of the zeroth generalized momentum and Gauss law, respectively.} This means that $|\mathcal{F}|=\infty^{2}$. Dropping the $\infty$ and focusing only on the exponents, we thus have
\begin{equation}\label{Eqn FFofEM}
    2c = 2(6) - 0 - 2(2) = 8 \implies c=4,
\end{equation}
in agreement with the previous heuristic arguments. We take this to be the right way to count DFF according to Penrose's prescription.

So, something like the proposal from Henneaux and Teitelboim \cite{Henneaux:1992ig} for counting DOFs of a theory can certainly be put to work in setting Penrose's notion of FF on firmer footing. Before we move on, however, we should note several important limitations of this counting method. First, as already mentioned, the method is only appropriate for field theories which admit Hamiltonian description. We've remarked previously that this leaves out an important sector of GR and potentially other field theories of physical interest. Second, the counting algorithm works only for \textit{classical} field theories in the sense of this section, and in particular, it remains unclear how to count FF of a quantum theory. This issue becomes of central importance once we introduce Penrose's counting arguments against string theory which itself is primarily a quantum theory. Presumably, one way to extend the counting algorithm to quantum theories would be to recognize that many quantum theories can be thought of as quantizations of classical field theories and then identify the DFF of the quantum theory with the DFF of the associated classical field theory. However, this is problematic for two reasons: (i) not all quantum theories can be thought of as quantizations of some classical field theory;\footnote{For instance topological quantum field theories which depend only on the topology of the underlying space and don't have a natural Lagrangian or Hamiltonian, or loop quantum gravity.}
 and (ii) existence of strong-weak coupling dualities can be taken to suggest that a single quantum theory has more than one classical counterpart (cf.\ \cite{Polchinski:2014mva}, which we discuss further below). If any such two `classical limits' differ in DFF, it would suggest that the DFF of the quantum theory is ill-defined.

Technical issues aside, one should be aware of one further limitation of the counting algorithm adumbrated above, which has to do with interpretation of constrained Hamiltonian systems and gauge theories. Following Belot \cite{Belot1998}, one can distinguish between three interpretive stances in relation to gauge theories: \textit{literal interpretation}, \textit{simply gauge-invariant interpretation}, and \textit{coarse-grained gauge-invariant interpretation}. According to the literal interpretation, the physical states are in one-to-one correspondence with points in phase space $T^*\mathbb{Q}$, whereas according to the simply gauge-invariant interpretation it is the \textit{gauge orbits} which stand in this one-to-one correspondence. Lastly, according to the coarse-grained gauge-invariant interpretation, the correspondence between gauge orbits and physical states is many-to-one. Generally, any of these interpretations will be available and their relative merits and demerits need to be compared before a final conclusion is reached.\footnote{Cf. the debate regarding the ontology of electromagnetism in light of the Aharonov-Bohm effect.} However, it seems that on the Henneaux-Teitelboim counting method, one is invariably committed to something like the simply gauge-invariant interpretation of gauge theories since the DOFs pertaining to gauge transformations are always subtracted. As a result, we don't believe that the counting algorithm can be said to correctly capture the physical DOFs of the literal interpretation. Moreover, on the literal interpretation, the dynamics will typically be indeterministic and so it remains unclear even how to correctly count the DFF of such a theory. One way to proceed would perhaps be to count DFF as previously but with the additional proviso that the DFF count correctly captures only the \textit{empirically significant} DOFs (all the interpretations will typically be empirically equivalent). In this way, one would be able to evade the interpretive question while retaining a well-defined DFF count. Perhaps there exists yet another way in which DFF of literally interpreted gauge theories could be counted: simply define $2c$ to be the dimension of the full, unconstrained phase space. Literal interpretations would then indeed come out with larger DFF count than gauge-invariant interpretations. In summary, there are at least two salient ways how to count DFF for literally interpreted gauge theories:
\begin{enumerate}
    \item Count only the `empirically significant DOFs' by defining the DFF to be the DFF of the gauge-invariantly interpreted theory. \label{count1}
    \item Count all the DOFs in which case the DFF is read off from the full phase space. \label{count2}
\end{enumerate} 
Once again, it should be noted that we are not aware of Penrose ever suggesting any such distinction or even discussing the role of interpretation in counting DFF. That said, we are slightly more inclined to regard \eqref{count2} as closer to Penrose's heuristic examples of \S\ref{Penrose on FF} which simply count all the admissible field configurations the space of which is obviously vastly larger for the literal interpretations.

Taking stock, the above survey of limitations of the counting algorithm makes us wonder whether DFF truly picks out a sufficiently robust and salient feature of physical theories. Moreover, in the next sections, we further argue that it cannot do the job that Penrose wants it to deliver in his counting arguments.

\subsection{Functional Freedom as a Criterion of Equivalence} \label{equivalence}
Recall that Penrose maintains that ``a theory cannot be equivalent to another [...] theory on which the initial space has a different number $D$ of dimensions," since ``the freedom in the $D$-space theory always vastly exceeds that in the [other] theory!'' \cite[p. 41]{FFF}. Therefore, we believe we are correct to read Penrose as proposing the following necessary condition on equivalence of theories:
\begin{quote}
    \textbf{Equivalence criterion (EC)}: Theories $\mathcal{T}_1$ and $\mathcal{T}_2$ are equivalent only if $\text{DFF}(\mathcal{T}_1) = \text{DFF}(\mathcal{T}_2)$.
\end{quote}
Here for simplicity we take $\mathcal{T}_1$ and $\mathcal{T}_2$ to be classical field theories bracketing for the moment the issue of how to extend the DFF count to quantum theories. If EC turned out to be sound, then this would certainly count towards the usefulness of FF and perhaps also towards its explanatory power. As a result, it would grant the concept of FF some legitimacy and support in becoming a new mathematical notion of quantity as we discussed in the previous section.

The notion of theoretical equivalence which was invoked informally by Penrose and which features in EC calls urgently for clarification. Let us therefore attempt to remove the point of confusion by offering some precisifications of the term. First of all, are we indeed concerned with empirical equivalence or perhaps with a stronger notion of complete physical equivalence? Consider the following precisification of EC:
\begin{quote}
    \textbf{$\text{EC}_1$}: Theories $\mathcal{T}_1$ and $\mathcal{T}_2$ are physically equivalent only if $\text{DFF}(\mathcal{T}_1) = \text{DFF}(\mathcal{T}_2)$.
\end{quote}
Physical equivalence is admittedly the tightest conceivable relationship between two physical theories: physically equivalent theories should be, after all, indistinguishable for the purposes of physics. We believe this grants $\text{EC}_1$ at least some credibility; however, in order to present convincing evidence in its favour---or a convincing counterexample for that matter---one would have to find two classical field theories which are uncontroversially physically equivalent and check whether they agree on their DFF counts. However, what exactly constitutes physical equivalence and the identity of scientific theories remains an unsettled point of discussion within the philosophical community. As a result, we find it difficult to reach a final verdict regarding $\text{EC}_1$ and in the absence of convincing counterexamples or supporting evidence, we would like to remain neutral regarding $\text{EC}_1$ pending further discovery of such examples and clarification of the notion of physical equivalence. This does not affect our subsequent discussion significantly since Penrose is more plausibly understood as appealing to either of the two principles which we next introduce.

Let's return to the topic of empirical equivalence. As far as this is concerned, philosophy of science operates with two version of this concept: \textit{weak empirical equivalence} and \textit{strong empirical equivalence}, which roughly translate to empirical equivalence on phenomena observed so far (or on phenomena belonging to some limited subset, e.g.~observed before some time $T$) and empirical equivalence on all possible phenomena, respectively. Could we precisify EC in either of the above senses? Let's consider:
\begin{quote}
    \textbf{$\text{EC}_2$}: Theories $\mathcal{T}_1$ and $\mathcal{T}_2$ are weakly empirically equivalent only if $\text{DFF}(\mathcal{T}_1) = \text{DFF}(\mathcal{T}_2)$.

    \textbf{$\text{EC}_3$}: Theories $\mathcal{T}_1$ and $\mathcal{T}_2$ are strongly empirically equivalent only if $\text{DFF}(\mathcal{T}_1) = \text{DFF}(\mathcal{T}_2)$.
\end{quote}
$\text{EC}_2$ is quite plainly false. As history seems to suggest, at least some instances of progress in physics have been accompanied by the discovery of new degrees of freedom. For example, the gravitational field in Newtonian gravity is fully determined and characterized by a single scalar potential $\phi$ whereas in general relativity a ten-component metric tensor is required. And yet, GR and Newtonian gravity are in the appropriate regime empirically equivalent; they are therefore weakly empirically equivalent.

We don't mean to argue that progress in physics \textit{must} be accompanied by a discovery of new degrees of freedom which increase FF, but we would like to make the point that discrepancy is to be expected either way.\footnote{Note that the it's also logically possible that the successor theory might not even be formalizable as a field theory, in which case Penrose's argument is not applicable in any straightforward way.} Furthermore, such discrepancy doesn't need to prevent the successor theory from being empirically adequate unless a clear link between the new degrees of freedom and empirical predictions is provided. Once such a link is provided, it still remains to be shown how the empirical predictions conflict with the ones of the predecessor theory; however, one doesn't seem entitled to make claims about weak empirical equivalence on the basis of DFF \textit{alone}.

This leaves $\text{EC}_3$ the only option on the table. 
Unfortunately for $\text{EC}_3$, we believe there are genuine counterexamples of strongly empirically equivalent theories which nevertheless disagree on DFF. One such counterexample is readily provided by the literal interpretations of gauge theories discussed in \S\ref{Sec Constraints}. The literal interpretation will generally be strongly empirically equivalent to the simply gauge-invariant and yet, if one counts DFF of the literal interpretations according to \eqref{count2}, then this will generally be different from the DFF of the simply gauge-invariant one. One could still maintain that the question of interpretation \textit{should not} make a difference to DFF counts and embrace \eqref{count1} as the correct counting method in which case the counterexample is blocked. Previously, we've expressed sympathies towards counting DFF according to \eqref{count2} as opposed to \eqref{count1} but we don't wish to be dogmatic in this regard. In fact, we concede that $\text{EC}_3$ for classical field theories could be defensible if \eqref{count1} is espoused instead; however, we note that this commits one to DFF being blind to the question of interpretation of gauge theories.

Having evaluated three distinct EC principles for classical field theories, we turn now to the realm of quantum theories. Since string theory itself is supposed to be a quantum theory of gravity, it is a quantum version of EC rather than the classical one which is effective in Penrose's counting arguments. What if we now allow $\mathcal{T}_1$ and $\mathcal{T}_2$ to be quantum theories? Pending a counting algorithm for quantum theories we adopt the naïve principle according to which the DFF of a quantum theory $\mathcal{T}_Q$ is the DFF of the classical field theory $\mathcal{T}$ associated to $\mathcal{T}_Q$ via quantization. Then $\text{EC}_2$ remains highly implausible for roughly similar reasons as before and moreover, new counterexamples $\text{EC}_3$ emerge.
For instance, what has become known as the worldline approach to QFT (see e.g.~\cite{schubert1997introduction}) is a blatant violation of $\text{EC}_3$.
In the worldline approach to QFT,
scattering amplitudes are no longer calculated from spacetime fields but rather from worldline fields by performing the path integral of worldline models.\footnote{See for instance the model introduced in the next section in equation \eqref{particle action}.} Remarkably, the scattering amplitudes obtained via the worldline methods agree with the ones obtained from second quantization of fields despite the fact that the functional freedoms of the two theories differ significantly! This is because the base manifold of the worldline models is one-dimensional and so the DFF involved would be something like $\infty^{c\infty^0}$ which is clearly much less than the DFF usually encountered in the empirically equivalent spacetime field theory. The worldline approach to QFT presents a particularly convincing problem case for $\text{EC}_3$ because it stands in direct analogy to perturbative string theory as we will see in the next section.\footnote{Relatedly, examples of exact holographic dualities such as the ones derived in \cite{costello2017holography,Eberhardt2020} also appear to contradict the quantum version of $\text{EC}_3$, for similar reasons.}


In the classical case, the defensible versions of EC are therefore (a) $\text{EC}_1$, the status of which is inconclusive given unsettled debates in the literature on theoretical equivalence, and (b) $\text{EC}_3$, provided that one espouses \eqref{count1} as the correct DFF counting method for gauge theories. In the quantum case, the situation is significantly less clear absent a convincing criterion for counting DFF. Nevertheless, we believe that only $\text{EC}_1$ is plausible in the quantum realm since $\text{EC}_3$ appears readily to be contradicted in the context of certain examples such as worldline QFT. We reject $\text{EC}_2$ in both classical and quantum cases.

\section{String Theory as a Field Theory} \label{string theory}

In this section, we outline the basics of perturbative string theory (PST) and string field theory (SFT), couching both in the field-theoretic language of \S\ref{FF}. For the sake of simplicity, we discuss here only the the bosonic string; however, it should be pointed out that realistic string models of quantum gravity necessarily will be superstring models which equip the theory with (worldsheet or target space) supersymmetry. For introduction to the bosonic string, see the canonical texts such as Polchinski \cite[ch.~1]{Polchinksi}, or Green-Schwartz-Witten \cite[chs.~1--3]{GSW}. A nice and elementary introduction to the bosonic string intended for philosophers is given by Huggett and Wüthrich \cite{HuggettWuthrich}.


The structure of the section is this. In \S\ref{sec:classicalstring}, we introduce the classical relativistic string. In \S\ref{sec:quantize}, we discuss quantization of the bosonic string. In \S\ref{sec:SFT}, we introduce string field theory.

\subsection{The Classical Relativistic String}\label{sec:classicalstring}
Discussions of perturbative string theory (PST) usually begin with a classical treatment of a relativistic string propagating in a flat Minkowski spacetime of arbitrary dimension $D$. Classically, a relativistic string in Minkowski spacetime is described by the `Nambu-Goto action'
\begin{equation} \label{NG action}
    S_{\text{NG}} = -\frac{1}{2\pi} \int_{\Sigma} d^2 \sigma \ \sqrt{-\det \Bigl(\frac{\partial X^\mu}{\partial\sigma^a} \frac{\partial X^\nu}{\partial\sigma^b} \eta_{\mu\nu} \Bigr)},
\end{equation}
where $\Sigma$ denotes the string worldsheet and the functions $X^\mu$ describe the embedding of the worldsheet into the Minkowski spacetime. Importantly, note that $S_{\text{NG}}$ is Poincaré invariant and is also invariant under worldsheet reparametrizations. One might recognize that the Nambu-Goto action in fact calculates the spacetime area of the string worldsheet which supplies it with a nice geometrical interpretation. Due to this geometrical interpretation, Poincaré invariance and reparametrization invariance shouldn't come as a surprise since this area is a coordinate-independent and parametrization-independent quantity. What makes $S_{\text{NG}}$ a natural choice for the string action is the analogy with the worldline action of a bosonic point particle which is proportional to the length of its the worldline. The action of the bosonic point particle is given by
\begin{equation} \label{particle action}
    S_{\text{BP}} = -m \int_{\gamma} \sqrt{-\eta_{\mu\nu} dX^\mu dX^\nu} = -m \int d\tau \sqrt{-\dot{X}^2}.
\end{equation}
Starting with a classical action such as (\ref{NG action}) or (\ref{particle action}), one can employ various methods of quantization in order to obtain a quantum theory associated with the original classical theory.
However, quantization of theories like (\ref{NG action}) is intractable due to the square roots appearing in the action. In string theory, one therefore resorts to the classically equivalent `Polyakov action',
\begin{equation} \label{Pol action}
    S_{\text{P}} = -\frac{1}{4\pi} \int_{\Sigma} d^2 \sigma \ \sqrt{-h} h^{ab} \partial_a X^\mu \partial_b X^\nu \eta_{\mu\nu},
\end{equation}
where one introduces the auxiliary worldsheet metric $h_{ab}$ which may be removed via its equation of motion to return (\ref{NG action}). Note that on top of the reparametrization and Poincare invariance of (\ref{NG action}), $S_{\text{P}}$ is also invariant under Weyl rescaling $h_{ab} \rightarrow e^{\omega(\tau,\sigma)} h_{ab}$, which is of crucial importance when it comes to quantizing the theory. It is this quantization of $S_{\text{P}}$ which we next describe.

\subsection{Quantization of the Bosonic String}\label{sec:quantize}

There are various ways in which one can quantize (\ref{Pol action}). To take here just one: one views the Polyakov action as describing a two-dimensional conformal field theory on the worldsheet and performs a path integral over the worldsheet fields which include the $X^\mu$ as well as the worldsheet metric $h_{ab}$ and a set of so-called \textit{ghost fields} which don't admit of any physical interpretation but the introduction of which is a necessary mathematical trick required by the reparametrization invariance of (\ref{Pol action}). From this point of view, the DFF of PST turns out to be something like $\infty^{a\infty^1}$, where $a$ is the number of degrees of freedom hidden in all the worldsheet fields.\footnote{
We are assuming (with Penrose) here and in what follows that the above discussions of counting FF in classical theories carry over to the context of quantum theories---our thanks to a reviewer for pushing us to be clear on this.}

The spectrum of the string (i.e., the Hilbert space of string states) can be found in the cohomology of the so-called BRST operator. The BRST operator is a nilpotent operator constructed from the ghost fields and generators of gauge symmetries of (\ref{Pol action}). We need not concern ourselves with the technical aspects of the BRST approach; however, let us describe briefly the structure of the open/closed string Hilbert space. It turns out that the Hilbert spaces of both open and closed bosonic string contain infinitely many states corresponding to particles of ever-increasing mass and tensorial rank. For example, the open string spectrum famously contains the scalar tachyon, a massless gauge-field reminiscent of the familiar Maxwell field, and further excited states of higher mass. The closed string spectrum too contains a tachyon state; however, the first excited state is no longer a vector field but rather a massless rank-two tensor field $f_{\mu\nu}$ which may be decomposed into three parts: an antisymmetric tensor $B_{\mu\nu}$, a traceless symmetric tensor $h_{\mu\nu}$, and the trace part $\Phi$, which are commonly referred to as the Kalb-Rammond field, the graviton, and the dilaton, respectively. In light of the presence of the $h_{\mu\nu}$ field in particular, a quantum description of the gravitational field is said to emerge naturally from the closed string spectrum.

Let's turn our attention back to the path integral over worldsheet fields mentioned above. The path-integral quantization is fraught with anomalies: failures of the classical symmetries to be preserved also at the quantum level. Imposing anomaly-freeness as a consistency condition of the entire process, one obtains further physically interesting conditions on the background fields in which the string propagates. First of all, anomaly-freeness requires one to set the critical dimension of the Minkowski background to $D=26$ (for the superstring, the critical dimension is $D=10$). Furthermore, the background metric is required to satisfy the Einstein field equations (EFEs) to first order in perturbation theory and corrections thereto in higher orders. More specifically, the dynamics of the background fields is dictated by the vanishing of beta functions (see e.g. \cite[ch.~3]{Polchinksi})
\begin{align} \label{beta f}
\begin{aligned}
    \beta_g &= \alpha' R_{\mu\nu} + 2 \alpha' \nabla_{\mu}\Phi \nabla_{\nu}\Phi + O(\alpha'^2) ,\\
    \beta_{\Phi} &= \frac{D-26}{6} - \frac{\alpha'}{2} \nabla^2 \Phi + \alpha' \nabla_{\mu}\Phi\nabla^{\mu}\Phi + O(\alpha'^2).
    \end{aligned}
\end{align}
Thus, familiar dynamics of the background fields emerges as a byproduct of the quantization process. For more on quantization of the bosonic string, see \cite[chs.~2--3]{GSW}.\footnote{For discussions of these issues in the philosophical literature, see \cite{Huggett2015-HUGDGR-3, Read2019-REAOMA}.}

There seems to be a common consensus in the string-theoretic community that the background fields themselves are not ontologically primitive objects but become identified with the so-called \textit{coherent string states} which are constructed from the string spectrum itself. Coherent states of quantum fields
are special `quasi-classical states' defined by the requirement that the number-phase uncertainty relations are saturated (see \cite[ch.~8]{Duncan2012}). A quick rundown of the standard string-theoretic argumentation regarding the stringy origin of the background fields is provided in \cite[ch.~5]{ReadBI}; a thorough analysis of the background field reduction may be found in \cite{HuggettWuthrich}. Whether the appeal to coherent states amounts to a legitimate and complete ontological reduction of the background fields still remains a moot point. For example, Read worries whether such a reduction presupposes an appeal to a spacetime theory beyond PST itself (a theory such as string field theory which we introduce shortly in \S\ref{sec:SFT}) \cite[ch.~5]{ReadBI}. We do indeed believe that such an appeal is required and we return to this problem again in the following section. Its significance will become clear as we progress through many of Penrose's criticisms; for now, we observe only that na\"{i}vely the theory of string background fields  seems to have KFF of the form $\infty^{b\infty^D}$, where $b$ comprises the degrees of freedom of all the background fields considered.\footnote{In the full theory, there are of course infinitely many such fields, as each excited state of the string gives rise to at least one such background field.}


At this point, we would like to suggest---as an aside---a potential connection between the discussion of background fields in PST and the literature on principle and constructive theories, this of course being a distinction going back to Einstein \cite{Einstein1919}.\footnote{For well-known philosophical discussion of this distinction, see \cite{Brown}.} While it would be nice if PST \textit{could} provide an ontological reduction of the background fields, the success of the theory itself doesn't seem to hinge on this particular project. String scattering amplitudes (which exhaust the empirical content of the theory) can still be calculated once the background is imposed `by hand', subject to all the relevant constraints such as the EFEs. The background fields in PST would thus simply be treated as primitive entities with behaviour constrained by imposing the condition of anomaly-free quantization. Such a refusal to provide an ontological reduction doesn't seem to hinder the \textit{predictive power} of PST; however, it does hinder its \textit{explanatory power}. If more variables are treated as primitive/imposed by hand in theory $\mathcal{T}_1$ as opposed to $\mathcal{T}_2$, then $\mathcal{T}_1$ would perhaps be regarded as \textit{less} explanatory than $\mathcal{T}_2$ which does offer a reduction.\footnote{This point is nicely exemplified in the relationship between thermodynamics and statistical mechanics. While thermodynamics treats all the thermodynamic variables as primitive, statistical mechanics \textit{explains} their origin via properties of statistical ensembles. As a result, one usually regards statistical mechanics as more explanatory.}

We suggest there to be a grain of analogy between this interpretative stance on PST and Einstein's distinction between principle and constructive theories. On this analogy, PST itself could be understood as a principle theory of scattering processes, while a spacetime theory such as string field theory (on which see the next subsection) would be its constructive counterpart. Recall that Einstein characterized principle theories as theories which take a set of basic, empirically grounded postulates and elevate them to the status of axioms which are subsequently employed in deductive reasoning about the phenomena which are thus explained by constraint but fall short of a constructive explanation. A constructive theory, according to Einstein, would then be one which provides explanation in terms of the microphysical underpinnings of the processes involved. Arguably, PST too proceeds from a postulate of some sort (although, of course, less directly connected to the empirical than those of Einstein in his 1905 article on special relativity \cite{Einstein1905}): one presupposes the existence of strings in background spacetime and proceeds to seek an anomaly-free Lorentz covariant quantum theory thereof. Whether the postulate is empirically well-grounded is perhaps questionable (see our above parenthetical); however, what in our opinion makes PST a principle(-like) theory is this axiomatico-deductive construction.\footnote{Cf. \cite{ALR}.} Furthermore, the dynamics of the background fields are explained \textit{by constraint} rather than by constructive microphysical underpinning, which is again seems characteristic of principle theories.

But if PST is indeed in this sense akin to a principle theory, does there exists a constructive counterpart governing the same phenomena? One constructive counterpart of PST could perhaps be found in string field theory, which we now introduce and interrogate.

\subsection{String Field Theory}\label{sec:SFT}
The moral of the previous discussion can be summarized as follows: perturbative string theory can be viewed as a two-dimensional conformal field theory which upon path integral quantization yields a recipe for calculation of S-matrix elements of various string scattering events as well as consistency conditions on the background in which the strings propagate. In the field theoretic framework of \S\ref{FF}, the KPMs of PST are thus picked out by the tuples $\langle \Sigma, X^1,..., X^D \rangle$.
There exists, however, an alternative and more general point of view in which string theory is formulated as a field theory on $D$-dimensional spacetime rather than the two-dimensional worldsheet: it is the open/closed bosonic string field theory (SFT). A recent and extensive introduction to both closed and open SFT can be found in \cite{Erbin}.\footnote{To be clear, SFT is just one approach to going beyond the framework of PST---others include e.g.\ M-theory and F-theory. We focus on SFT since it is amenable to an at least somewhat tolerably clear counting of FF, in diagloue with Penrose's criticisms of string theory.}

The fundamental objects of SFT are the so-called open/closed string fields $\Phi$ and $\Psi$, respectively, which live on a background  spacetime manifold $M$.
From the field theoretic perspective, the KPMs of SFT are thus tuples $\langle M,  \Phi, \Psi \rangle$, but the simple appearance is highly misleading since both $\Phi$ and $\Psi$ are both highly structured objects involving infinitely many degrees of freedom. Recall from our previous discussion of the string spectrum that excitations of the open string give rise to a tachyon, a massless gauge boson,
and infinitely many more massive particles of higher spin labelled by a discrete quantum number $\alpha$ and a continuous momentum label $k$. In SFT, one incorporates all these excitations into a single mathematical object: the open string field
\begin{equation}\label{SF expansion}
    |\Phi\rangle = \sum_{\alpha} \int \frac{d^Dk}{(2\pi)^D} \psi_{\alpha}(k) |k,\alpha\rangle.
\end{equation}
Fourier transforming the components $\psi_{\alpha}(k)$ then yields a spacetime field corresponding to each label $\alpha$. Furthermore, substituting (\ref{SF expansion}) into an appropriate action functional yields a total spacetime action for all the fields in the spectrum of the open string.\footnote{For the free open SFT, this action takes the form $S_{\text{SFT}} = \frac{1}{2} \langle\Phi|Q_{B}|\Phi\rangle$.} According to Erbin, one should therefore view SFT as an ordinary QFT with the following features: the amplitudes agree with the worldsheet amplitudes, but the number of fields is infinite, and their interaction non-local, see \cite[p.~26]{Erbin}. To put it mildly, a theory like SFT is thus very difficult to grasp from the point of functional freedom. Because the KPMs are in fact infinite tuples, the KFF of the full SFT must be something like $\infty^{\infty \infty^D}$.

We would like to leave aside the technical details of the SFT project and instead briefly fill in the gaps in the story of ontological reduction of the background fields outlined in \S\ref{sec:quantize}. Recall that background fields are supposed to be ontologically reduced coherent string states spread across spacetime. In SFT one can accommodate this requirement straightforwardly since the string field is by construction a quantum field on spacetime (in fact, infinitely many quantum fields). Thus, in our view, background fields in string theory should be thought of as \textit{coherent states of the string field}. We believe that this appeal to SFT completes the ontological reduction outlined in \cite{HuggettWuthrich} and thus clarifies the origin of background fields in string theory. 


\section{Penrose's Arguments} \label{arguments}

Having explicated what Penrose means by functional freedom, 
we proceed now to assess what we take to be Penrose's main arguments against string theory. The arguments assessed in \S\ref{counting} are those which involve explicitly the counting FF in string theory; as a result, our above criticisms of the concept of FF undermine at least partially these arguments. The objections to string theory assessed in \S\S\ref{excitation}--\ref{initial data} don't depend in any crucial way on the concept of FF and so warrant independent assessment.

\subsection{Counting Functional Freedom} \label{counting}

There are three main `FF counting arguments' against string theory scattered across \cite{RR} and \cite{FFF}:
\begin{enumerate}
    \item The `classical physics argument'.
    \item The `heterotic string argument'.
    \item The `AdS-CFT argument'.
\end{enumerate}
In the following three subsections, we present and assess each of these in turn.

\subsubsection{The Classical Physics Argument} \label{classical arg}
In \emph{Fashion, Faith and Fantasy}, Penrose says the following regarding bosonic string theory formulated as a two-dimensional conformal field theory on the worldsheet:
\begin{quote}
    In my attempts to get to grips with the various developments in string theory, there has been an additional point of potential confusion for me, particularly when trying to understand the issues of functional freedom. [...] It is particularly difficult to assess the functional freedom that is involved in a physical theory unless one has a clear idea of its actual space-time dimensionality. To be more explicit about this issue, let me turn to one of the particularly appealing aspects of the early string ideas [...] [I]n the early days of string theory, the subject was sometimes viewed from the point of view of a 2-dimensional conformal field theory [...] 
    This would lead us to a picture in which the functional freedom had the form $\infty^{a\infty^1}$ for some positive number $a$. How are we to square this with the far larger functional freedom $\infty^{b\infty^3}$ that is required for ordinary physics? [...]
    The point I am making here is certainly not that functional freedom might in some sense be ill defined or irrelevant. The point is, however, that for a theory formulated in a way dependent on things like power series coefficients or mode analysis, it may not be at all easy to ascertain what the functional freedom actually is [...]. \cite[pp.~62--3]{FFF}
\end{quote}
We read Penrose as expressing in the above passage a twofold worry. He observes the discrepancy in DFF between worldsheet PST on the one hand and what he calls `ordinary physics' on the other and concludes that this is somehow problematic. We take this to be an implicit appeal to some form of EC, presumably $\text{EC}_2$. Secondly, he observes that the full functional freedom of worldsheet string theory might not be easily ascertainable (or perhaps manifest) from the formulation of the theory. The two worries are importantly of entirely different philosophical character: the former is metaphysical while the other is epistemic.\footnote{An anonymous reviewer has suggested to us that Penrose could mean here that the notion of FF is ill-defined. However, this cannot be what he intends, since he rules this out explicitly in the above passage.}

We don't see why the epistemic worry should count as an objection to string theory. After all, it seems to point at a limitation of the theorist rather than at a problem within the theory. Admittedly, easily ascertainable functional freedom could be seen as a super-empirical virtue of a theory, but certainly absence of this quality can't be seen as a vice.

With regards to the metaphysical worry, we would like to make clear that we don't believe that either PST or SFT should be obliged somehow to reproduce the FF of `ordinary physics' as Penrose insists; further, the implicit appeal to $\text{EC}_2$ is illegitimate. Setting aside the question of what `ordinary physics' means in this context and, relatedly, the question of what exact FF should these theories aim to reproduce, we recall that based on our previous discussion $\text{EC}_2$ in its quantum or classical guise is not a sound principle! Thus, we don't see any problem in the DFF discrepancy between the worldsheet theory and ordinary physics. (Indeed, that one can derive the emprirical predictions of `ordinary physics' from less FF could be marketed to be an advantage; moreover, the whole issue is presumably moot when one moves to SFT, which has more FF anyway.\footnote{Our thanks to Noel Swanson for discussion on this point.}) The former may still plausibly reproduce predictions of the latter in the appropriate experimental regime. Morever, we don't regard it as very plausible for Penrose to be invoking $\text{EC}_1$ or $\text{EC}_3$ in this argument since string theory is a successor theory to classical physics and so physical equivalence or strong empirical equivalence is not desirable.

Penrose seems to appeal to EC also in the following passage; however, this time the problem seems to be excessive DFF of the background fields:
\begin{quote}
    Accordingly, deep questions are raised concerning the physical relevance of quantum theories such as supra-dimensional string theories, for which the number of spatial dimensions is greater than three we directly perceive. What happens to the floods of excessive degrees of freedom that now become available to the system, by virtue of the huge functional freedom that is potentially available in the extra spatial dimensions? Is it plausible that these vast numbers of degrees of freedom can be kept hidden away and prevented from completely dominating the physics of the world in such schemes? \cite[pp.~41--2]{FFF}
\end{quote}
But what exactly does it mean for degrees of freedom to `dominate the physics of the world'? The most plausible reading of this phrase seems to be that the mismatch in DFF would threaten the \textit{empirical adequacy} of theories like PST or SFT which in an appropriate experimental regime should aim to reproduce predictions of general relativity and quantum field theory. As such, the claim then seems to be once again appealing to $\text{EC}_2$ or $\text{EC}_3$, both of which we have called into question in \S\ref{equivalence}. 

Notwithstanding our above criticisms, we share sympathies with Penrose insofar as he can be read as asking for a clear geometrical picture of string theory and in particular, clarification of the status of background fields in PST---a topic we have investigated in \S\ref{sec:SFT}. We agree that unless one appeals to (e.g.)~SFT and coherent states of the string field, the move from the worldsheet formalism to the background fields remains obscure and in those circumstances, asking how is it possible to make a leap from the worldsheet formalism to the spacetime theory of background fields would be entirely apposite.\footnote{However, we  still maintain that FF isn't relevant to these considerations.}

\subsubsection{The Heterotic Sting}
Another FF-counting objection concerns one of the superstring theories: the heterotic string (for an introduction, see e.g.~\cite{Polchinksi}). Once again, Penrose seems to be calling for a consistent geometrical picture:
\begin{quote}
    The strange feature about heterotic string theory is that it appears to behave simultaneously as a theory in 26 space-time dimensions and in 10 space-time dimensions (the latter with accompanying supersymmetry), depending upon whether we are concerned with left-moving or right-moving excitations of the string
    [...] This dimensional conflict would seem to cause us problems if we are to try to work out the functional freedom involved
    [...] 
    I find it very difficult to form a consistent picture of what is going on here, and I have never even seen these geometrical issues properly discussed. \cite[pp.~63--5]{FFF}
\end{quote}
In principle, we don't see why our discussion of coherent string field states wouldn't extend also to the superstring case. One must remember that the number of spacetime dimensions, should confusion arise, is in the worldsheet formalism precisely the number of bosonic $X^{\mu}$ fields. Erbin sums up the situation regarding the heterotic string as follows:
\begin{quote}
    The ghost super-CFT is characterized by anti-commuting ghosts $(b,c)$ (left-moving) and $(\Bar{b},\Bar{c})$ (right-moving) with central charge $c=(-26,26)$, associated to diffeomorphisms, and by commuting ghosts $(\beta, \gamma)$ with central charge $c=(11,0)$, associated to local supersymmetry. As a consequence the matter SCFT must have a cetral charge $c=(15,26)$. If spacetime has $D$ non-compact dimensions, then the matter CFT is made of:
    \begin{itemize}
        \item a free theory of $D$ scalars $X^{\mu}$ and $D$ left-moving fremions $\psi^{\mu}$ $(\mu = 0, \dots, D-1)$ such that $c_{\text{free}} = D$;
        \item an internal theory with $c_{\text{int}} = 15 - 3D/2$ and $\Tilde{c}_{\text{int}} = 26 -D$.
    \end{itemize}
    The critical dimension is reached when $c_{\text{int}} = 0$ which corresponds to $D=10$. \cite[p.~245]{Erbin}
\end{quote}
He even identifies the same problem as Penrose when he writes that ``since the critical dimension of the two sectors do not match, one needs to get rid of the additional dimensions of the right-moving sector'' \cite[p.~19]{Erbin}. In spite of this, the heterotic string field theory seems to be unproblematically formulated on $D$-dimensional spacetime. Would this be a consistent geometrical picture for Penrose?

At the same time, we repeat that we remain sceptical with regards to the relevance FF considerations. Our inability to ascertain FF shouldn't be taken as a vice of the theory.

\subsubsection{The AdS-CFT Correspondence}

Finally, we would like to draw attention to FF-related concerns which Penrose raises in the context of the AdS-CFT correspondence (first developed in \cite{Maldacena:1997re}; for a philosophical introduction, see e.g.\ \cite{Rickles2013-RICADA-3}). Put simply, Penrose sees a problem in the mismatch in dimensionality (and hence FF) between the AdS and CFT sides of the duality and consequently takes this to indicate that the theories cannot in fact be equivalent. Consider a particular instantiation of the AdS-CFT correspondence and let $M$ be the ten-dimensional product $\text{AdS}_5 \times S^5$ and $E$ be the conformal infinity of $\text{AdS}_5$. Penrose then notes that
\begin{quote}
    [h]ere there is no chance of appealing to the type of `quantum-energy' argument put forward in §31.10 [see \S\ref{excitation}] for explaining away the gross discrepancy between the functional freedom of an ordinary field on $M$, namely $\infty^{M\infty^9}$ and an ordinary field on [$E$], namely [$\infty^{E\infty^3}$]. Since the extra dimensions of $M$ are in no way ‘small’---being of cosmological scale---the flood of additional degrees of freedom, from the fields’ dependence on the $S^5$ part of $M$, would spoil any possibility of an agreement between the two field theories. The same would apply to ordinary QFTs on $M$ and [$E$], since one-particle states are themselves described simply by ‘ordinary fields’ [...]. The only chance of the holographic principle being actually true for these spaces is for the QFTs under consideration to be far from ‘ordinary’. \cite[p.~921]{RR}
\end{quote}
On pain of repetition, we once again fail to see how FF considerations should bear on the issue of theory equivalence. Our argumentation is therefore perfectly analogous to the one delivered in \S\ref{classical arg}. The only thing which, indeed, we might wish to add here is that although it is true \emph{classically} that a bulk spacetime cannot be determined by its asymptotic boundary data, this is not true quantum mechanically already at the semi-classical level: see \cite{Wall, Raju2, Raju1}. If these results are correct and generalise, then there is perhaps room to argue that the DFF of bulk and boundary theories in cases of holographic duality are in fact equal, in which case the objections raised by Penrose in this argument are moot.\footnote{We're grateful to Henrique Gomes for discussion on this final point.}\textsuperscript{,}\footnote{Another related observation here is that, if one cashes out dualities \emph{à la} Polchinski \cite{Polchinski:2014mva}---one quantum theory having multiple classical limits---then it is far from obvious that dual theories need agree on FF. Our thanks to Jacob Barandes for raising this point with us.}




\subsection{Excitation of Extra Degrees of Freedom} \label{excitation}

It is now time to introduce and assess two central arguments presented in \cite{RR} and \cite{FFF} which regard, respectively, `quantum' and `classical' considerations.
In this subsection we take up the quantum argument only to connect it with the classical argument in the following subsection.

First, let us give a brief primer on compactification in string theory. We have  seen above that vanishing of the relevant (`Weyl') anomaly enforces a `critical dimension' on string-theoretic spacetimes which turns out to be 26 for the bosonic string and 10 for the superstring. In order to recover four-dimensional phenomenology, string theorists have studied models in which the `additional' 22 or 6 dimensions are compactified into a manifold of small size. `Four-dimensional phenomenology' in the above refers, foremost of all, to the phenomenology of particles which we have observed so far in particle accelerators. For the superstring, this compact manifold is further required to be a special kind of complex three-manifold, a Calabi-Yau threefold, so that the string-theoretic spacetime takes the product form $M \times X$, where $M$ is the four-dimensional Minkowski spacetime and $X$ is the Calabi-Yau. The number of possible Calabi-Yau compactifications is enormous ($10^{500}$ is sometimes quoted) and this leads to a proliferation of string theoretic particle models since different compactifications generically yield different string spectra. This has led some to worry about predictive and explanatory power of string theory giving rise to the so-called \textit{landscape problem} of string theory, on which see \cite{Read2021-REATLA-2}.

Before we articulate the quantum argument, let's consider one motivating example. Following Penrose, we analyze the Klein-Gordon theory of a single scalar field $\phi$ on a spacetime with a single compactified dimension. Specifically, we consider the product form $M \times S^1$ where $M$ is the familiar four-dimensional Minkowski spacetime and $S^1$ is a circle of small radius $\rho$. The Klein-Gordon equation on $M \times S^1$ then takes the form
\begin{equation}
    \left(\Box - \frac{1}{\rho^2} \frac{\partial^2}{\partial \theta^2} + m^2 \right) \phi = 0,
\end{equation}
where $\theta$ denotes a coordinate along the $S^1$-compactified dimension and $\Box$ is the d'Alembertian in $M$. One may then separate out the modes associated with $S^1$ by taking $\phi = e^{in\theta} \psi$ so that the Klein-Gordon equation reduces to
\begin{equation}
\left( \Box + \frac{n^2}{\rho^2} + m^2 \right) \psi = 0.
\end{equation}
On standard particle physics reasoning, one can then interpret the $\frac{n^2}{\rho^2}$ term in the above as effectively contributing to the mass of the particles described by the scalar field. Furthermore, if the radius is small compared to the length scale of $m$, that is if $\rho \ll m^{-1}$, the effective mass of the particle would be so large that it would be virtually undetectable by current particle accelerators. The reason for this is that energy-mass equivalence would prohibit creation of such particles since the local particle collisions occur at much smaller COM energies in current particle accelerators.

It is exactly this piece of particle physics reasoning (as employed by string theorists) which Penrose attacks with his quantum argument. As we will see shortly, Penrose seems to believe that such heavy particles \textit{could} be created given the conditions present in our universe. But before we analyze his claims, let us rephrase the above motivating example in string-theoretic terms.

Recall again that string-theoretic spacetimes take the product form $M \times X$ where $X$ is a Calabi-Yau. Following Penrose  \cite{Penrose_2002}, one may privilege a time axis on $M$ and study the dynamics of fields on $M \times X$ as an initial value problem of modes on $\mathbb{R}^3 \times X$. Due to the compactness of $X$, the modes associated with $X$ would form a discrete family in analogy to the Klein-Gordon example. Penrose then notes that
\begin{quote}
    [...] the energy of excitation of a [$X$-mode] is expected to be very large because of the very minute scale of [$X$] itself. A dynamical `standing wave' on [$X$] would have a tiny wavelength, comparable to the Planck distance of $~10^{-33}$ cm, and would therefore have something like a Planck frequency of $~10^{-43}$ seconds. The energy required to excite such a mode would be of the general order of a Planck energy, [...], which is nearly twenty orders of magnitude larger than the largest energies involved in ordinary particle interactions. It is accordingly argued that the modes that affect [$X$'s] geometry will remain unexcited, in all particle-physics processes that are of relevance to physical actions available today. \cite[p.~191]{Penrose_2002}
\end{quote}
The problem with this kind of reasoning in string theory, Penrose contends, is that the Planck-order energies are in fact available in our universe which means that the $X$-modes should get excited arguably leading to the ``unleashing of floods of extra-dimensional degrees of freedom that are potentially there by virtue of the freedom that is in the Planck-scale geometry'' and to ``devastating effects on the macroscopic dynamics'' \cite[p.~74]{FFF} presumably in the form of a proliferation of otherwise-unobserved heavy particles. Penrose thus asks:
\begin{quote}
    [A]re the positive energy (Planck-scale) modes of vibration of the six extra dimensions immune from excitation? Although the Planck energy is indeed very large when compared with normal particle-physics energies, it is still not that big an energy, being comparable with the energy released in the explosion of about one tonne of TNT. There is, of course, enormously more energy than this available in the known universe. For example, the energy received from the Sun by the Earth in one second is some $10^8$ times larger! On energy terms alone, that would be far more than sufficient to excite the [$X$] space for the entire universe! \cite[p.~903]{RR}
\end{quote}
In particular, passages such as \cite[pp.~72--3]{FFF} indicate that Penrose rejects the idea that such an energy would need to be delivered in a localized manner by an energetic particle:
\begin{quote}
    But we must bear in mind that the picture that the string theorists are presenting is one in which the space-time [...] would be taken as a product space $M \times X$ [...]. If the extra dimensions themselves were to be excited, the relevant `excited mode' [...] of the space-time would be exhibited as our higher-dimensional space-time having the form $M \times X'$, where $X'$ is the perturbed (i.e.~`excited') system of extra dimensions.
    (Of course, we have to think of $X'$ as being, in some sense, a `quantum space', rather than a classical one, but this does not seriously affect the discussion.)
    A point that I am making here is that in perturbing $M \times X$ to $M \times X'$, we have perturbed the \textit{entire universe} [...] so that when we are thinking of the energy required to effect this mode of perturbation as being `large' we must think of this in the context of the universe as a whole. It seems to me to be quite unreasonable to demand that the injection of this quantum of energy be necessarily effected by some fairly localized high-energy particle. \cite[p.~72]{FFF}
\end{quote}

Let us now take a step back to assess this criticism. Crucial in the above passage seems to be the notion of exciting a degree of freedom. Unfortunately, what exactly this means (especially in the context of string theory) is unclear to us. All the reader is offered is an analogy concerning the quantized energy levels of a hydrogen atom which in non-relativistic QM become excited by incoming photons \cite[p.~70]{FFF}. While such a picture might serve as useful heuristic or approximate description of atomic processes, the extension of this concept to \textit{fields} or \textit{modes} requires further justification. In the context of atomic transitions, it may be argued that the appeal to abrupt transitions between energy levels is in fact just a proxy description justified by the underlying theory of light and matter. So what could be the underlying microphysical story for background string theory? Unless such a story is provided, one may doubt whether Penrose engages in sound physical reasoning in the above passages.

Perhaps one attempt to cash out this story would be via an appeal to SFT. String fields $\Phi$ and $\Psi$ are fields on spacetime and their various components may become excited as ordinary quantum fields. We note though that Penrose doesn't mention SFT explicitly in the relevant passages and in fact discussion of SFT is completely absent from his exposition to string theory. Thus, whether the picture Penrose has in mind in the above passages is something along the lines of SFT or perhaps something completely different is a question which we cannot answer definitively. We believe there are reasons to think that Penrose in fact didn't have the SFT picture in mind because it doesn't quite square with the rest of the quantum argument. For if excitations of the string field around the background coherent state were to occur, they would surely need to be facilitated by localized means rather than TNT explosions, something which Penrose explicitly rejects.\footnote{For simplicity, we neglect the fact that SFT actually involves non-local interactions.}

Let us be a little more explicit about this issue. The difference between localized and non-localized energy injections is presumably in \textit{energy density}, or more precisely, in the values of the stress-energy tensor. The energy of a macroscopic event such as a TNT explosion is spread across much larger spacetime volume than the energy of a localized energetic particle and as a result, the energy density of a TNT explosion will generally be much smaller than the density of an energetic particle despite the \textit{total energy} being roughly similar. The crucial point is this: from a QFT perspective (which we can take to coincide with the SFT perspective), it is \textit{energy density} which matters for field excitations to occur, not the total energy. Heuristically speaking, this is because smaller values of stress-energy tensor at a point correspond to smaller field values at that point and any kind of point-wise interaction between fields would also in turn out to be smaller.

Independently of the above, there are further reasons to be sceptical of the claims about TNT explosions and energy delivered by the sun. First, we don't see why the the problem should be any more relevant to string theory than it is to ordinary quantum field theory. If TNT explosions or sunlight were sufficient to create particles in string theory, surely they would also be sufficient to create particles in the Standard Model and yet, Penrose doesn't seem to be worried about the same problem in the Standard Model. Why not? Second, we note that TNT explosions are not the typical kind of regime in which predictions of particle physics ordinarily get tested. In principle, there's nothing wrong with that but since this is the case, more should be said about how exactly predictions of string theory would get violated during, e.g., a TNT explosion. After all, string theory and the Standard Model are tested primarily in scattering experiments, so what predictions do they make for a TNT explosion anyway?

Let us summarize the discussion so far. We began by pointing out the difficulty in understanding what Penrose means by excitation of degrees of freedom, that is unless one adopts a field theory perspective such as SFT or QFT. However, in field theories energy needs to be delivered in a localized manner for excitation of quantum fields to occur and consequently, claims about TNT explosions and sunlight exciting degrees of freedom don't make much sense. Furthermore, we see neither how these considerations apply to string theory any more than they do to ordinary quantum field theory, nor what their precise empirical significance is in fact supposed to be.


Later on, Penrose also mentions the tension between his hypostatized, abrupt quantum transitions and a commonsense notion of locality. He writes:
\begin{quote}
    [...] the ground state of $X$ is, by its very nature, of necessity \textit{not} localized at any particular place in our ordinary space-time $M$, being supposed to be omnipresent, permeating the structure of space-time throughout the entire universe. The geometrical quantum state of $X$ is supposed to influence the detailed physics that is going on in the most remote galaxy, just as much as here on Earth. The string theorist's argument that Planck-scale energy would be far too great, in relation to what is available, to be able to excite $X$ seems to me to be inappropriate on various counts. Not only are such energies amply available through non-localized means [...], but if we were to imagine that $X$ were actually to be converted to an excited state $X'$ by such a particle transition (perhaps owing to some advanced technology making a Planck-energy particle accelerator), [...] this would be clearly absurd, as we could not expect the physics on the Andromeda Galaxy to be instantly changed by such an event here on Earth!
    We should be thinking more in terms of a much milder event in the vicinity of the Earth propagating outwards with the speed of light. Such things would be described much more plausibly by nonlinear classical equations, rather than abrupt quantum transitions. \cite[pp.~74--5]{FFF}
\end{quote}
We find the above passage puzzling because the tension with locality certainly doesn't stem from string theory but rather from the quantum transitions themselves. Brief inspection of (\ref{beta f}) shows that dynamics of background fields in string theory are codified in covariant equations which enforce relativistic causality. As a result, we fail to see how the above passage constitutes an objection against string theory, in spite of being phrased as such. More plausibly, we believe that it highlights the inappropriateness and limitations of using quantum transitions to describe the system.

One should also be aware of a potential point of confusion. When Penrose says that the ``\textit{entire universe}'' is excited, thereby affecting physics in e.g.~the Andromeda galaxy, this would seem to presuppose that all copies of $X$ distributed over $M$ will get excited in the same way. But how does this follow? Surely, $M \times X$ is a product space and any excitation of $X$ would happen locally in $M \times X$! That is, at a particular point $p \in M$ one excites the $X$ space but this \textit{does not} automatically translate to excitations of $X$ in the Andromeda galaxy.



In summary, we believe there to be much confusing terminology and unclear argumentation involved in the quantum argument. The notion of excitaiton of degrees of freedom is insufficiently cashed out and although it could perhaps be made sensible by an appeal to SFT, we don't think this squares well with what Penrose says.

\subsection{Classical Instability of Extra Degrees of Freedom} \label{singularity}
Let us now investigate the argument from classical instability of string theoretic spacetimes which may be found in both \cite{RR} and \cite{FFF}. The argument relies on a singularity theorem in general relativity proven by Hawking and Penrose  \cite{Hawking1970}; subsequently, Penrose applies it to compactified spacetimes of the kind encountered in string theory.
The assumption behind Penrose's reasoning seems to be that background fields in string theory are to an excellent approximation described by the EFEs derived from the vanishing of the beta functions, as we discussed in detail in \S\ref{sec:quantize}. However, note that this is indeed only an approximation and that the full dynamics of the background fields is governed by the vanishing of the \textit{entire} power series (\ref{beta f}), not merely the first couple of terms in string constant $\alpha'$. Setting this issue aside for a moment, let us delve deeper into the singularity theorem.

The statement of the singularity theorem is roughly the following. Once again, consider a compactified string-theoretic spacetime $M \times X$, where $X$ is a small Calabi-Yau threefold and $M$ is the Minkowski spacetime. Since $M = \mathbb{R}^3 \times \mathbb{R}$, we can regroup the product terms and define $Z := \mathbb{R} \times X$. We now impose perturbations on an initial-value surface in $Z$ and study the propagation of this perturbation into the future. The theorem of Hawking and Penrose then states that the evolution will result in a singularity provided that:
\begin{enumerate}
    \item The dimension of $Z$ is $n\geq3$ and $Z$ contains a compact $(n-1)$ dimensional hypersurface without closed timelike curves.
    \item The Einstein tensor satisfies the  \textit{strong energy condition} (SEC).\footnote{For a statement and general discussion of energy conditions, see \cite{Curiel_2017}.}
\end{enumerate}
We note with Penrose that a singularity by itself doesn't necessarily imply infinite curvature but simply a failure to continue the evolution of the initial data surface into the future. Such a failure might arise due to infinite curvatures of the subspace $Z$ but need not necessarily do so. On this point, Penrose observes that ``[a]lthough there are alternative things that can in principle happen in exceptional cases, it is to be expected that the general reason for the impossibility of continuing the evolution is that curvatures indeed do diverge" \cite[p.~81]{FFF}. What seems to be at risk again is the empirical adequacy of string theory since we don't seem to observe such violent instabilities in our universe. We quote:
\begin{quote}
    What this singularity theorem appears to be telling us is that so long as the perturbations of the extra dimensions can be treated classically---as indeed appears to be a reasonable thing to do, as a clear conclusion of our earlier considerations [...]---then we must expect a violent instability in the 6 extra spatial dimensions, in which they crumple up and approach a singular state. [...]
    Whatever the extra dimensions might crumple to, the observed physics is not likely to be other than drastically affected. This is hardly a comfortable picture of the 10-dimensional space-time that string theorists have been proposing for our universe. \cite[pp.~81--2]{FFF}
\end{quote}
Although the conclusion drawn from the singularity theorem indeed sounds serious, we aver that one should be suspicious the applicability of the theorem in the context of string theory. In particular, we question the applicability of the SEC to this context. Penrose seems to believe that SEC is ``certainly satisfied'' \cite[p.~81]{FFF} in string theory because it follows directly from the vacuum EFEs which Penrose assumes to hold on $M \times X$. But this supposition is twofold inaccurate. Recall that the full dynamics of background fields in string theory is \textit{not} given by the EFEs but rather by the vanishing of the entire beta functions (\ref{beta f}) which include infinitely many more terms and that EFEs emerge only as a low-order approximation, so an appeal to the EFEs in justifying the SEC can be at best only approximate.\footnote{This, furthermore, gives rise to an ambiguity in what SEC actually means. Once higher order corrections to EFE are taken into account, one has to be careful about whether the \textit{geometric} or \textit{physical} SEC is to be imposed because the two are no longer straightforwardly equivalent via the EFE (see \cite{Curiel_2017} for a review of energy conditions). Since Penrose talks about the \textit{Einstein tensor} satisfying the SEC, we will assume that he in fact means the geometric SEC, which states that $R_{\mu\nu}X^\mu X^\nu \geq 0$ for all timelike $X^\mu$.}
However, even granting Penrose the low-order approximation of the full dynamics, the \textit{vacuum} EFEs are most certainly not an accurate model for string theoretic backgrounds. This is because the spectrum of the string contains not only the graviton, but also infinitely many more massive particles which all jointly contribute to the total energy tensor featuring on the right-hand side of the equations. To assume a vacuum would be to neglect degrees of freedom which, importantly, \textit{cannot} be neglected in string theory.

Answering the following thus appears to be crucial for the classical-instability argument to go through: does the full stress-energy tensor of background fields in string theory satisfy the physical SEC? In a recent article \cite{Parikh_2015}, Parikh and van der Schaar prove that the geometric null energy condition (NEC) is satisfied in string theory. However, we note that while geometric SEC implies geometric NEC, the converse doesn't hold so the result unfortunately doesn't help the argument to get off the ground.


Penrose himself raises the concern that the SEC might fail to obtain once higher-order terms in $\alpha'$ are taken into account. Yet he proceeds with his argument justifying this step by an appeal to practices of string theorists:
\begin{quote}
    Another point of relevance here is that the strong energy condition being assumed here, although automatically satisfied by [$R_{\mu\nu}=0$], certainly cannot be guaranteed if we are to consider what happens with the higher-order terms in the power series in $\alpha'$, referred to above. Yet, most current considerations of string theory seem to operate at the level where these higher-order terms in $\alpha'$ are ignored [...] \cite[p.~81]{FFF} 
\end{quote}
As we argued above, this move is insufficient to get the argument off the ground since neglecting the matter fields is a highly unrealistic scenario in string theory. In \cite[p.~907]{RR}, Penrose comments further on the extendibility of his theorem to the cases when the full dynamics is taken into account:
\begin{quote}
    We should also take note of the fact [...] that $(1+9)$-dimensional Ricci flatness is not precisely the requirement that string theory demands. We recall that Ricci flatness is regarded merely as an excellent approximation to that requirement, coming about when terms higher than the lowest order in the string constant $\alpha'$ are ignored. Maybe the ‘exact’ requirement, involving all orders in the string constant $\alpha'$, could evade the above singularity theorem. However, if this requirement provides us with a condition on the Ricci tensor for which the usual local energy-positivity demands are satisfied [...], then the singularity theorem would still apply. On the other hand, violations of such local energy conditions can certainly occur in QFT [...], so these issues are far from conclusive. \cite[p.~907]{RR}
\end{quote}
And so it seems that the status of the classical instability argument too is far from conclusive. The next term in the expansion of the gravitational beta function (\ref{beta f}) is proportional to $R_{\mu\kappa\lambda\tau}R_{\nu}^{\kappa\lambda\tau}$ \cite[p.~178]{GSW} but higher order terms are notoriously difficult to calculate. In light of the above discussion, it would be interesting to see whether the theory enriched by this additional term satisfies the geometric/physical SEC. However, as we have argued, that SEC obtains hasn't been demonstrated convincingly  even in the first-order approximation given by the EFE. So, in our view, the classical instability argument remains inconclusive.

\subsection{The Problem of Initial Data} \label{initial data}
The final argument from Penrose against string theory which we consider in this article once again concerns the dynamics of the background fields which, as we have repeated many times now, are described by the system of equations (\ref{beta f}) containing infinitely many terms of ever-increasing order. This feature of the system is allegedly problematic:
\begin{quote}
    More serious, to my mind, is the fact that the full requirement, involving all orders in the string constant $\alpha'$, is actually and infinite system of differential equations of unbounded differential order. Accordingly, the data that would be needed on an initial 9-surface would involve derivatives of all orders in the field quantities (rather than just thee first or second derivatives that are needed in ordinary field theories). The number of parameters per point needed on the 9-surface is then infinite, so we get a functional freedom greater than $\infty^{M\infty^9}$, for any positive integer $M$. This would seem to make the problem of excessive functional freedom even worse than before! I am not aware of any serious discussion of the mathematical form of this full requirement, and of what kind of initial data might be appropriate for it. \cite[p.~907]{RR}
\end{quote}
Having dispatched FF-counting arguments in \S\ref{counting}, we don't dwell on the FF-related point in the above passage but discuss rather the alleged problem with infinitely many terms and unbounded order. We don't regard the infinitude of terms and their unbounded differential order as inherently problematic for string theory because, in principle, the Cauchy problem could be still well-defined and the system's solutions existent and well-behaved. That is, as long as convergence is secured. It is precisely this silent assumption which Penrose calls into question when he notes that
\begin{quote}
    [b]ecause of the extreme smallness of $\alpha'$, however, the higher-order terms are usually ignored in specific versions of string theory that are put forward (though the validity of doing this is in some doubt, as there is no information concerning the convergence or ultimate behaviour of the series [...]). \cite[p.~78]{FFF}
\end{quote}
With regards to convergence we believe that Penrose's criticism is perfectly justified and to this day, we are unaware of any discussion of the convergence problem in the literature. In fact, beta functions generically don't converge which seems to make the problem yet more pressing.\footnote{Note that this would also seem to make Penrose's argument here applicable to any theory expressed as a power series, e.g.\ many QFTs. (Our thanks to the anonymous reviewer for stressing this to us.)} Unfortunately, calculation of higher-order terms in (\ref{beta f}) has proven computationally challenging which hinders progress with regards to this question.

\section{Conclusion} \label{conclusion}

Let's take stock. Penrose, in his \emph{The Road to Reality} \cite{RR} and his \emph{Faith, Fashion and Fantasy}, criticises string theory on numerous grounds, often---but not always---invoking the notion of functional freedom. In this article, we've seen that the notion of functional freedom is difficult to formalize mathematically and does not deliver in the respects that Penrose's arguments require.
Moreover, Penrose's criticism of string theory are at worst specious or at best inconclusive.

All this being said, we should not end on too negative a note: Penrose should be lauded for his serious and critical engagement with the string theory research programme. In fact, we \emph{agree} that the extent to which string theory can go unquestioned in contemporary theoretical physics is problematic (cf.\ \cite{Smolin}); physics is surely in a healthier place for dissenting voices such as his.

\section*{Acknowledgements}

First and foremost, we are very grateful to Roger Penrose for generous exchanges on the topic of functional freedom. We are also very grateful to Jacob Barandes, Enrico Cinti,  Henrique Gomes, Sebastian de Haro, Nick Huggett, Ondřej Hulík, Brian Pitts, and Noel Swanson for helpful discussions. We also thank the anonymous reviewers for valuable feedback.


\bibliographystyle{plain}
 \bibliography{refs}

\end{document}